\documentclass[aps,pra,reprint,superscriptaddress,showkeys,floatfix,raggedbottom]{revtex4-2}
\usepackage[T1]{fontenc}
\usepackage{mathptmx}
\usepackage{microtype}
\usepackage{amsmath,amssymb}
\usepackage{graphicx}
\usepackage{booktabs}
\usepackage{needspace}
\usepackage{silence}
\usepackage[hidelinks]{hyperref}
\usepackage{url}
\begin{document}

\title{Composition-Dependent Fragmentation and Dielectric Response of
Hydrogen-Bond Networks in Water--Ethanol Electrolytes}
\author{Vishwajit Rajesh}
\affiliation{Department of Chemical Engineering, National Institute of Technology Tiruchirappalli}
\author{Shiladitya Sengupta}
\affiliation{Department of Physics, Indian Institute of Technology Roorkee}
\author{Nandlal Pingua}
\email{nandlal@nitt.edu}
\affiliation{Department of Chemical Engineering, National Institute of Technology Tiruchirappalli}

\begin{abstract}
We examine hydrogen-bond network connectivity and solvent dielectric response
in water--ethanol--NaCl mixtures near 1~mol\,kg\(^{-1}\). The largest
water-only component falls sharply between solvent ethanol mole fractions
0.3511 and 0.5197, while a large mixed-solvent component persists at higher
ethanol content. A not-a-knot cubic-spline estimate places the steepest decline
near \(x^*=0.4759\) on the ten-composition grid (\(0.4335\) on the nine-point
grid; see Supplemental Material), and the largest water cluster stops wrapping
across the periodic cell somewhere in the same interval. Donor-site-normalized
counts show that the dominant direction of cross-species donation reverses with
composition. The finite-field solvent response decreases from
\(47.89\pm0.70\) to \(18.87\pm0.26\) at \(E_z=0.005\ \mathrm{V/\text{\AA}}\),
and a second field gives consistent values. A structural contact-ion
association index rises about 63-fold across the nine compositions and tracks
the inverse dielectric response; the two quantities come from separate
trajectories, so we report only that they co-vary. The network crossover is a
finite-cell feature, not an infinite-system critical point, and the
finite-field response is not taken to be a zero-field permittivity.
\end{abstract}

\maketitle

\section{Introduction}

Water--ethanol mixtures show composition-dependent hydrogen bonding and
nonideal structure. Diffraction and simulation studies have followed how these
networks change with composition and temperature, and graph-theoretical
simulations relate alcohol-aggregate morphology and water structure to
miscibility across alcohol--water systems.
\citep{pothoczki-2021,choi-parameswaran-2020}
Local coordination and collective connectivity are separate properties:
counting bonded neighbors says nothing about whether a species forms a large
or system-spanning component.

Molecular-dynamics analyses have examined hydrogen-bonded pairs in
ethanol--water mixtures, whereas component and wrapping analyses address
collective connectivity.
\citep{gereben-pusztai-2015,pothoczki-2021} We state an explicit
donor--acceptor geometry and report directed counts, component size, and
periodic wrapping as separate descriptors.

The broad connectivity behavior is established in salt-free binary mixtures.
Gereben and Pusztai found water-only 3D percolation at up to 40~mol\% ethanol
(placing the water-only threshold between 40 and 60~mol\%), while the total
mixed network remained percolating through their sampled range up to
80~mol\% ethanol. Pothoczki et al. place the water-subnetwork percolation
transition near \(x_{\mathrm{EtOH}}=0.5\) at room temperature.
\citep{gereben-pusztai-2017,pothoczki-2021} Pair-level studies of ethanol--water
hydrogen bonds and a combined computational/ATR study of water--ethanol
cross-association are related precedents.
\citep{gereben-pusztai-2015,drougkas-2024} Di Fonte et al. introduced a
directed-graph connectivity descriptor distinguishing donor (``broadcaster'')
from acceptor (``receiver'') roles and applied it to water and aqueous NaCl
solutions. Their analysis found ionic effects on water--water hydrogen-bond
connectivity largely confined to the first hydration shell.
\citep{di-fonte-2026} We therefore do not present water-network fragmentation,
mixed-network percolation, or graph-based hydrogen-bond analysis as new concepts. Our focus
is their composition-resolved co-analysis with finite-cell wrapping,
finite-field solvent response, and structural ion association in
NaCl-containing water--ethanol mixtures.

Dielectric spectra add a dynamic view: simulations link ethanol--water
dielectric response to collective reorganization of the hydrogen-bond network
as well as to faster bond formation and breakage.
\citep{cardona-sweatman-lue-2018} The finite-field solvent response computed
here differs from both a frequency-resolved spectrum and a zero-field static
permittivity.

Salt perturbs these networks in a composition-dependent way. NMR measurements
show that NaCl disrupts water structure in 20~vol\% ethanol, whereas at
60~vol\% ethanol lower NaCl concentrations strengthen water--ethanol
structure and promote proton exchange.\citep{nose-hojo-ueda-2004}
Simulations of MgCl\(_2\) in water--ethanol likewise find composition-dependent
stabilities of contact and solvent-assisted ion pairs, though the different
cation and conditions make this only adjacent context for NaCl.
\citep{chatterjee-dixit-tembe-2013} A uniform salt effect cannot be assumed,
so we analyze network topology, solvent polarization, and ion contacts
separately.

We study water-only and mixed-solvent networks in water--ethanol--NaCl
mixtures using directed bond counts, largest-component fractions, and periodic
wrapping of the largest water cluster. Directed counts give donor-site-normalized
bond populations; wrapping tests periodic spanning. Temporal persistence of
solvent hydrogen-bond and Na--Cl contact edges is shown in Supplemental
Fig.~S5. We also estimate the solvent-only finite-field response at two field
strengths and compare its composition trend with a salt-free experimental
reference and a structural contact-ion index. These electrostatic quantities
are kept separate from network topology. The network-size and water-cluster
wrapping results are summarized in Figure~\ref{fig:network-connectivity}.
Together, these analyses provide a cross-scale view of composition-dependent
structure and response.

\begin{figure*}[!t]
\centering
\begin{minipage}[t]{0.48\textwidth}
\centering
\textbf{(a)}\par\smallskip
\includegraphics[width=\linewidth]{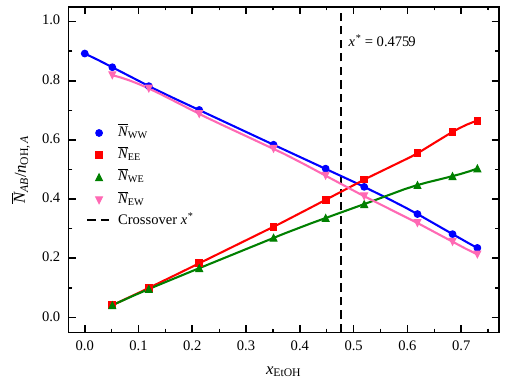}
\end{minipage}\hfill
\begin{minipage}[t]{0.48\textwidth}
\centering
\textbf{(b)}\par\smallskip
\includegraphics[width=\linewidth]{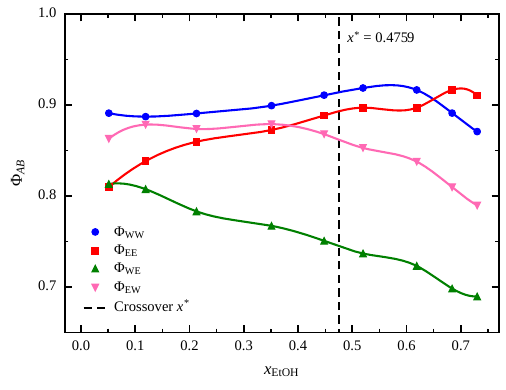}
\end{minipage}
\caption{Composition dependence of directed hydrogen-bond populations.
(a) Bond counts normalized by donor O--H sites. (b) The same counts divided
additionally by the salt-free bulk mole fraction of the acceptor species; this
is a bulk-abundance correction, not an intrinsic acceptor probability. In both
panels, the dashed line marks the spline estimate \(x^*=0.4759\) of the
steepest decline in the water-only largest-component fraction (see
Supplemental Material), shown for reference. Points are composition means,
connecting lines guide the eye, and no uncertainty bars are shown.}
\label{fig:directed-comparison}
\end{figure*}

\section{Methods}

\subsection{Systems and molecular models}

The systems contained 3,697 water molecules and 0, 200, 500, 1,000, 2,000,
4,000, 6,000, 8,000, or 10,000 ethanol molecules. An additional system with
3,000 ethanol molecules served as an intermediate-composition validation point
for the network analysis. Each system contained equal numbers of
\(\mathrm{Na^+}\) and \(\mathrm{Cl^-}\) ions, with the salt loading set to
approximately 1~mol\,kg\(^{-1}\) on the total water--ethanol solvent-mass
basis. Water was modeled with standard TIP4P/2005 parameterization
\citep{abascal-vega-2005}
having oxygen-site Lennard--Jones parameters
\(\sigma_{\mathrm O}=3.1589~\text{\AA}\) and\linebreak[0]
\(\epsilon_{\mathrm O}=0.1852~\mathrm{kcal\,mol^{-1}}\), and a virtual charge
site displaced 0.1546~\AA{} along the H--O--H bisector. Ethanol was modeled
with OPLS-AA and ions with the Madrid-2019 scaled-charge model
\citep{jorgensen-1996,zeron-2019}.
Compression and NPT relaxation, and production-run stability diagnostics, are
summarized in the Supplemental Material (Figs.~S6--S11 and Tables~S5--S6).

Ethanol composition is the salt-free solvent mole fraction,
\(x_{\mathrm{EtOH}}=N_{\mathrm{EtOH}}/(N_{\mathrm{water}}+N_{\mathrm{EtOH}})\).
The nine-composition production series spans \(x_{\mathrm{EtOH}}=0\) to
0.7301. The 3,000-ethanol system has \(x_{\mathrm{EtOH}}=0.447962\); it was
used to validate the network interpolation and was left out of the dielectric
and ion-association series. Rounding the number of NaCl pairs to an integer
gave actual molalities between 0.9995 and 1.0060~mol\,kg\(^{-1}\) on the
combined solvent-mass basis. Each mixture was equilibrated at 300~K and 1~atm,
and the final NPT box volume was kept for fixed-volume production. Across the
main composition series, the relaxed density decreases from approximately
1.04~g\,cm\(^{-3}\) to 0.86~g\,cm\(^{-3}\).

\begin{figure*}[!t]
\centering
\begin{minipage}[t]{0.49\textwidth}
\centering
\textbf{(a) Largest-component fractions}\par\smallskip
\includegraphics[width=\linewidth]{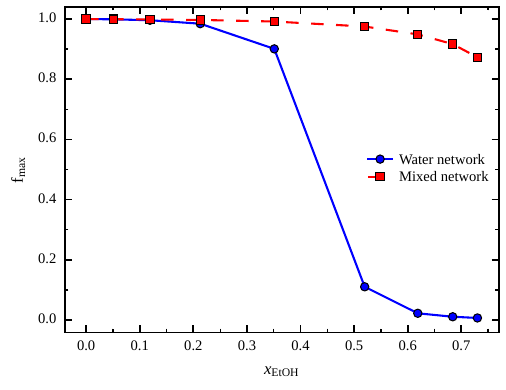}
\end{minipage}\hfill
\begin{minipage}[t]{0.49\textwidth}
\centering
\textbf{(b) Water-cluster wrapping probability}\par\smallskip
\includegraphics[width=\linewidth]{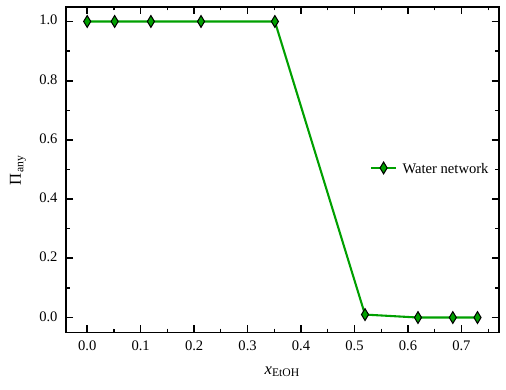}
\end{minipage}
\caption{Water-network size and periodic spanning across composition.
(a) Largest-component fractions for water-only and mixed water--ethanol
graphs; the large mixed component persists after the water-only network
fragments. (b) Empirical wrapping probability of the largest water cluster.
The validation composition at \(x_{\mathrm{EtOH}}=0.447962\), with empirical
wrapping probability 0.8503, is omitted because wrapping is intermittent.
The probability is 0.00998 (5/501 frames) at 0.5197 and zero at 0.6187--0.7301;
the sampled change occurs between 0.3511 and 0.5197.}
\label{fig:network-connectivity}
\end{figure*}

\Needspace{9\baselineskip}
\subsection{Hydrogen-bond graphs and directed counts}

Each water or ethanol molecule was a node. Following the directed
donor-to-acceptor graph convention of Di Fonte et al.~\citep{di-fonte-2026}, we
accepted a directed hydrogen bond from donor molecule \(D\) to acceptor
molecule \(A\) when the donor--acceptor oxygen distance was within its
first-shell cutoff and the angle
\[
 \theta_{\mathrm{HO_dO}}=\angle(\mathrm{H_D-O_D\cdots O_A})
\]
was at most \(35^\circ\). The angle vertex is at the donor oxygen, so a
linear \(\mathrm{O_D-H_D\cdots O_A}\) geometry has angle zero. The
water--water distance cutoff was fixed at 3.35~\AA{}; cutoffs involving
ethanol were set at the corresponding first-shell minima. Atom-level
geometries were mapped to parent molecules, and duplicate molecular edges
were collapsed for undirected graph analysis. Directional counts were taken
before this deduplication.

In \(N_{AB}\), the first subscript is the donor and the second the acceptor,
with \(A,B\in\{\mathrm{W},\mathrm{E}\}\). Counts were normalized by available
donor O--H sites: two per water molecule and one per ethanol molecule. To
compare acceptor classes despite their different bulk abundance, we also
computed
\begin{equation}
 \Phi_{AB}=\frac{\overline{N}_{AB}/n_{\mathrm{OH},A}}{x_B},
 \qquad
 n_{\mathrm{OH},A}=\begin{cases}
 2N_{\mathrm{water}},& A=\mathrm{W},\\
 N_{\mathrm{EtOH}},& A=\mathrm{E},
 \end{cases}
 \label{eq:opportunity-index}
\end{equation}
where \(x_B=N_B/(N_{\mathrm{water}}+N_{\mathrm{EtOH}})\) is the
salt-free bulk mole fraction of the acceptor. This normalization combines the
directed donor/acceptor role separation of Di Fonte et al. with a
bulk-composition correction analogous in spirit to the H-bond-number excess
parameter used by Pothoczki et al.~\citep{pothoczki-2021,di-fonte-2026}. It
references bulk abundance only; it does not account for local encounter
frequency and is not an intrinsic acceptor probability.

\subsection{Components and periodic wrapping}

Water-only graphs contain water--water hydrogen-bond edges. Mixed-solvent
graphs contain water--water, ethanol--ethanol, and water--ethanol edges. For
each frame, connected components were identified and the largest-component
fraction computed as \(f_{\max}=S_{\max}/N\), where \(S_{\max}\) is the number
of molecules in the largest connected component. Here, \(N\) is the number of
water molecules for the water-only graph and the total number of solvent
molecules for the mixed graph.

Wrapping was evaluated separately from component size. Integer
periodic-image shifts were assigned to water-network edges and propagated
along graph paths, and a nonzero winding inconsistency marked a wrapped
cluster. The empirical probability that the largest water cluster wrapped in
at least one direction was
\begin{equation}
 \Pi_{\mathrm{any}}=\frac{1}{N_{\mathrm{frames}}}
 \sum_t W_{\mathrm{any}}(t),
 \label{eq:wrapping-probability}
\end{equation}
where \(W_{\mathrm{any}}(t)\) is one for a wrapped cluster and zero
otherwise. Wrapping of the mixed network was not analyzed.

\subsection{Finite-field solvent dielectric response}

Field-driven trajectories were run at 300~K in NVT at each composition's
final NPT-equilibrated box volume (the production-box volume of
Table~S5). A uniform field was applied along \(z\) at \(E_z=0.005\) and
\(0.007\ \mathrm{V/\text{\AA}}\), with both runs lasting 5~ns. The first
0.3~ns was discarded from means and uncertainty estimates. Standard errors
were computed from the retained dipole samples with the Geyer
initial-monotone-sequence autocorrelation estimator.\citep{geyer-1992} Augmented Dickey--Fuller
tests\citep{said-dickey-1984} over 0.5--5.0~ns reject the unit-root null at all nine compositions
for both fields (all \(p<1.1\times10^{-8}\); see Supplemental
Tables~S3--S4); this does not prove strict stationarity.

The solvent polarization includes neutral water and ethanol molecular dipoles
and excludes ionic charge displacements, which can reflect sustained ionic
current rather than bounded solvent polarization.\citep{cox-sprik-2019}
The fixed-volume finite-field estimator was
\begin{equation}
 M_{z,\mathrm{solv}}=\sum_{m\in\mathrm{water}}\mu_{m,z}
 +\sum_{m\in\mathrm{ethanol}}\mu_{m,z},
 \qquad
 \epsilon_s^{\mathrm{est}}=1+
 \frac{\langle M_{z,\mathrm{solv}}\rangle}{\epsilon_0 V_0 E_z},
 \label{eq:dielectric-estimator}
\end{equation}
where \(V_0\) is the fixed production-box volume of each composition (the
final NPT volume) and \(\epsilon_0\) is the vacuum permittivity. We call this
a finite-field solvent response and do not identify it with a zero-field
static permittivity, which would require evidence for linear response.

\subsection{Structural contact-ion association index}

To compare with the dielectric response, we use the dimensionless structural
index
\begin{equation}
 A_{\mathrm{CIP}}=\frac{f_{\mathrm{CIP}}}{f_{\mathrm{mono}}^2},
 \label{eq:cip-index}
\end{equation}
where \(f_{\mathrm{CIP}}\) is the fraction of ions with at least one Na--Cl
contact and \(f_{\mathrm{mono}}\) is the fraction with none. The
composition-specific first-shell Na--Cl contact boundaries span
3.35--3.48~\AA{}. Ions in larger contact-connected aggregates count toward
\(f_{\mathrm{CIP}}\), so the index is not restricted to isolated dimers, and it
is not a thermodynamic equilibrium constant. Ion-cluster histograms were
accumulated in separate, field-free 5-ns NVT trajectories with 1-ps sampling.
Differencing successive cumulative histograms gives a 1-ps series for the
contact fraction, and Geyer autocorrelation standard errors are used where
they converged. For the 6,000- and 10,000-ethanol systems, single-trajectory
uncertainties did not converge, so the final standard errors use four
independent trajectories (the original 5-ns run and three 2-ns replicas). The
association and dielectric data therefore come from unmatched trajectories and
are compared only across composition.

\begin{figure*}[!t]
\centering
\begin{minipage}[t]{0.48\textwidth}
\centering
\textbf{(a)}\par\smallskip
\includegraphics[width=\linewidth]{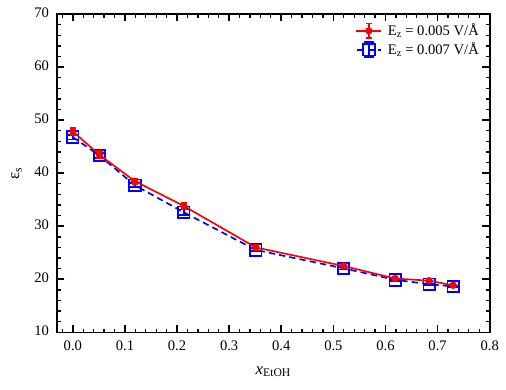}
\end{minipage}\hfill
\begin{minipage}[t]{0.48\textwidth}
\centering
\textbf{(b)}\par\smallskip
\includegraphics[width=\linewidth]{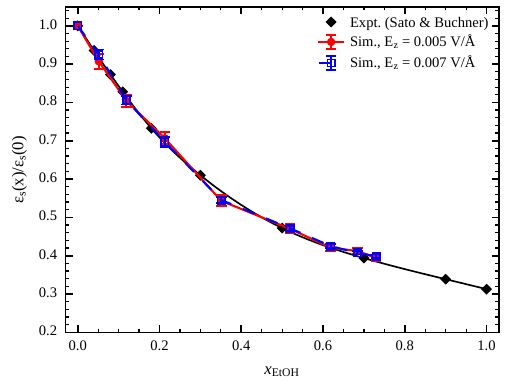}
\end{minipage}
\caption{Composition dependence of the solvent dielectric response.
(a) Finite-field estimates at \(E_z=0.005\) and
\(0.007\ \mathrm{V/\text{\AA}}\), with reported SEMs; the fields agree
within 2.1\(\sigma\) across the sampled compositions, with the largest
difference at \(x_{\mathrm{EtOH}}=0.6839\) (2.10\(\sigma\)). (b) Each
series normalized to its own zero-ethanol value
[\(\varepsilon_s(0)=47.89\pm0.70\) at \(0.005\ \mathrm{V/\text{\AA}}\)
and \(46.82\pm0.44\) at \(0.007\ \mathrm{V/\text{\AA}}\)]. The experimental
reference is salt-free water at 298.15~K; the electrolyte simulations are at
300~K. The normalization compares composition trends only, not absolute
permittivities or an isolated salt effect.}
\label{fig:dielectric-comparison}
\end{figure*}

\section{Results}

The donor-site-normalized cross-species counts change ordering across the
composition series. At \(x_{\mathrm{EtOH}}=0.0513\), ethanol-to-water
donation is 0.8186 bonds per ethanol O--H site, compared with 0.0417
water-to-ethanol bonds per water O--H site. At \(x_{\mathrm{EtOH}}=0.7301\),
water-to-ethanol donation is 0.5034 per water O--H site, compared with 0.2131
ethanol-to-water donation per ethanol O--H site. The raw ratio
\(N_{\mathrm{WE}}/N_{\mathrm{EW}}\), however, stays between 1.72 and 1.88, so
the change in normalized ordering comes from the changing donor populations.

Water self-donation per water O--H site decreases from 0.8917 in the
water-only system to 0.2351 at \(x_{\mathrm{EtOH}}=0.7301\), while ethanol
self-donation increases from 0.0416 at \(x_{\mathrm{EtOH}}=0.0513\) to
0.6648 at 0.7301. The two water-ending donor-normalized curves are similar
at both ends of the series, whereas the ethanol-ending curves separate at
high ethanol content. These are donor-normalized counts, not acceptor-preference
probabilities. Both trends are shown in Figure~\ref{fig:directed-comparison}(a,b).

After bulk-acceptor-fraction correction, the ethanol-donor index is higher
for water acceptors at \(x_{\mathrm{EtOH}}=0.0513\)
(0.8629 versus 0.8097). At \(x_{\mathrm{EtOH}}=0.4480\), the ordering
reverses: the index is higher for ethanol acceptors (0.8886 versus 0.8680),
a difference of 0.0206. At 0.7301 the corresponding values are 0.7896 and
0.9105. For water donors, the water-acceptor index remains higher across the
series, including 0.8708 versus 0.6895 at 0.7301. Because this correction
uses bulk abundance rather than local encounters, it reveals neither an
intrinsic acceptor preference nor which bonds sustain connectivity.

\subsection{Water-network fragmentation and wrapping}

The largest water-only component fraction declines from 0.998994 at
\(x_{\mathrm{EtOH}}=0\) to 0.900042 at 0.3511, then drops to
0.111014 at 0.5197, 0.023156 at 0.6187, and 0.007672 at
0.7301. The largest mixed-solvent component, in contrast, still
contains 0.974105 of solvent molecules at 0.5197, 0.948104
at 0.6187, and 0.872113 at 0.7301. A large mixed component thus
persists after the water-only graph fragments, although its size alone does
not show that the mixed network wraps or that ethanol-only edges form a
continuous path.

The 3,000-ethanol validation composition is
\(x_{\mathrm{EtOH}}=0.447962\), where the largest water-component fraction is
0.5804, consistent with the fragmentation profile. Including this point in
the cubic-spline series places the steepest decline at \(x^*=0.4759\); the
original-grid estimate is 0.4335. The interpolation and its angular-cutoff
sensitivity are reported in the Supplemental Material. The fitted location is a
finite-cell descriptor, not a directly sampled transition.

The largest water cluster wraps in every frame from $x_{\mathrm{EtOH}}=0$ through 0.3511, in 0.00998 of frames at 0.5197, and in none from 0.6187 to 0.7301. At the validation composition (0.447962) it wraps intermittently, with probability 0.8503 (omitted from the plot). The sampled values carry no uncertainty estimate, and the loss of wrapping brackets the crossover between 0.3511 and 0.5197 (Fig.~\ref{fig:network-connectivity}).

\begin{figure*}[!t]
\centering
\begin{minipage}[t]{0.48\textwidth}
\centering
\textbf{(a)}\par\smallskip
\includegraphics[width=\linewidth]{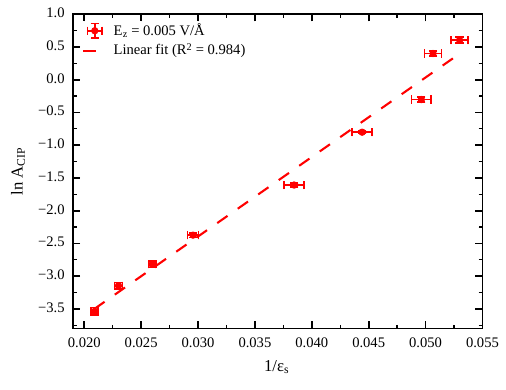}
\end{minipage}\hfill
\begin{minipage}[t]{0.48\textwidth}
\centering
\textbf{(b)}\par\smallskip
\includegraphics[width=\linewidth]{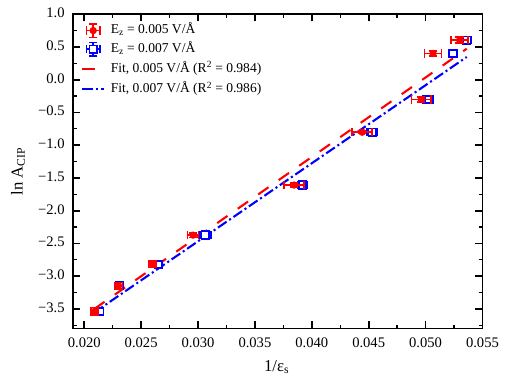}
\end{minipage}
\caption{Relationship between the structural index \(A_{\mathrm{CIP}}\) and
inverse finite-field solvent response across composition.
(a) Primary-field data at \(E_z=0.005\ \mathrm{V/\text{\AA}}\), with
horizontal SEMs in inverse dielectric response and vertical SEMs in
\(\ln A_{\mathrm{CIP}}\), with the unweighted linear fit. (b) Fits using
both tested field strengths with the same field-free association data and
the same vertical SEMs. The similarity of the fits shows consistency across
fields, not causality. The dielectric and association measurements come from
separate trajectories.}
\label{fig:association-dielectric}
\end{figure*}

\subsection{Finite-field dielectric response and field comparison}

At the primary field, the estimated solvent response decreases monotonically
from \(47.89\pm0.70\) at zero ethanol to \(18.87\pm0.26\) at
\(x_{\mathrm{EtOH}}=0.7301\); all systems contain NaCl. The independent
\(0.007\ \mathrm{V/\text{\AA}}\) series follows the same trend, and the
two field estimates agree within 2.1\(\sigma\) across all nine compositions, with
the largest standardized difference (2.10\(\sigma\)) at
\(x_{\mathrm{EtOH}}=0.6839\). This is a marginal difference at the
reported precision; it does not establish the zero-field limit or
linear response. Complete estimates are in the Supplemental Material and are
compared in Figure~\ref{fig:dielectric-comparison}(a).

\subsection{Normalized comparison with salt-free mixture data}

To compare composition trends, each dielectric series was normalized to its
own zero-ethanol value. The simulation reference is the zero-ethanol NaCl
electrolyte, whereas Sato and Buchner's experimental reference is salt-free
water at 298.15~K; the simulations are at 300~K.
\citep{sato-buchner-2004} The primary simulation ratio falls to 0.9075 at
\(x_{\mathrm{EtOH}}=0.0513\), 0.5431 at 0.3511, and 0.3940 at 0.7301.
The two simulation fields track each other and follow the experimental
composition trend over the sampled range. Salt content, baseline,
temperature, and observable all differ between simulation and experiment, so
this comparison gives context but cannot isolate an ion-induced dielectric
decrement. The normalized series are compared in
Figure~\ref{fig:dielectric-comparison}(b).

\subsection{Structural association and inverse dielectric response}

The structural index \(A_{\mathrm{CIP}}\) rises about 63-fold across the
nine compositions, including the zero-ethanol electrolyte. Unweighted
linear fits of \(\ln A_{\mathrm{CIP}}\) against inverse dielectric response
give
\begin{equation}
\begin{aligned}
 \ln A_{\mathrm{CIP}}&=(121.36\pm5.90)(1/\epsilon_s^{(0.005)})\\
 &\quad -(6.03\pm0.23),\quad R^2=0.984,\\[0.3em]
 \ln A_{\mathrm{CIP}}&=(119.12\pm5.40)(1/\epsilon_s^{(0.007)})\\
 &\quad -(6.04\pm0.22),\quad R^2=0.986.
\end{aligned}
 \label{eq:association-dielectric-fits}
\end{equation}
The power-law fits give exponents of \(4.20\pm0.31\) and \(4.19\pm0.31\),
with \(R^2=0.963\) at 0.005 and \(R^2=0.963\) at 0.007~V/\AA{}.
The exponential-in-inverse-response fits are better: their RMS residuals
are 0.187 and 0.175, versus 0.282 and 0.284 for the power laws, with
\(\Delta\mathrm{AIC}=7.44\) and 8.79 favoring the exponential fits.
Both fits use the same field-free association data.
Because the two observables came from separate trajectories, and because
changing composition also alters local solvation and contact geometry, the
fits show that the quantities co-vary; they do not establish a screening
mechanism or an equilibrium association constant. Both fits are shown in
Figure~\ref{fig:association-dielectric}(a,b). An additional analysis on data from the same trajectory is required.

\section{Discussion}

Locally, water self-donation decreases as ethanol self-donation rises, and
the donor-normalized cross-species counts reverse ordering. The raw ratio
stays at 1.72--1.88, so part of the reversal reflects changing donor
populations rather than a shift in acceptor preference. Counts also do not
identify which edges are necessary for spanning.

At the graph scale, the water-only largest component collapses while a large
mixed component persists. Ethanol therefore appears to stay in connected mixed
structures after water--water fragmentation, but without a wrapping analysis
of the mixed network we cannot say whether these structures span the cell.
Together with the water wrapping results, the data point to finite-cell
reorganization rather than a thermodynamic critical point.

The solvent-edge endpoint-overlap curves in Supplemental Fig.~S5 all fall
below 0.011 by 1,000~ps, which suggests that solvent connectivity survives
edge turnover. Ion-contact edges appear separately in that figure and are not
hydrogen bonds. Endpoint overlap allows edges to break and re-form between
observations and has no block-level uncertainty, so it is neither a full edge
lifetime nor evidence of dynamical slowing.

The finite-field solvent response decreases with ethanol, and the two fields
agree within about 2.1\(\sigma\), though this does not fix the zero-field
limit. The association index \(A_{\mathrm{CIP}}\) rises about 63-fold and
tracks inverse response. Since the dielectric and ion data come from separate
trajectories and composition changes local solvation and contact geometry,
the link remains a co-variation.

The observables measure different things: donor counts give bond populations,
components and wrapping give collective connectivity, edge overlap tracks
endpoint retention, and the finite-field estimator measures solvent-dipole
response.

Distinguishing mechanisms will require wrapping analysis of the mixed network,
finite-size and finer-composition studies, local-encounter-conditioned bond
statistics, matched polarization and contact sampling, and salt-free controls.

The study is limited by its coarse composition grid and single cell size.
Wrapping outcomes are sampled zeros and ones without block-resolved
uncertainty, and the validation point does not narrow the reported bracket.
Directed counts have no uncertainty bars, and the bulk-abundance correction
does not condition on local encounters.
\section{Conclusion}

As ethanol content increases, the water-only hydrogen-bond network fragments
in these finite cells. The spline estimate of steepest decline in the
water-only largest-component fraction is \(x^{*}=0.4335\) on the nine-point
grid and \(x^{*}=0.4759\) including the validation composition (see
Supplemental Material), and the loss of water-cluster wrapping falls between
0.3511 and 0.5197. A large mixed-solvent component
persists at higher ethanol fractions; whether it wraps was not tested.
Directed donation shifts from ethanol-to-water at low ethanol content to
water-to-ethanol at high ethanol content, as water self-donation declines
and ethanol self-donation rises. The solvent-only finite-field dielectric
estimate falls from 47.89 to 18.87 at the primary field (and from 46.82 to
18.65 at the second field), giving consistent values. Ion association rises
strongly across the nine compositions and tracks inverse dielectric
response, though the data come from separate trajectories and show
co-variation only. None of these measures, alone or together, establishes a
thermodynamic percolation transition or a causal mechanism.
\bibliography{references}
\end{document}


\maketitle
\renewcommand{\thefigure}{S\arabic{figure}}
\renewcommand{\thetable}{S\arabic{table}}
\renewcommand{\theequation}{S\arabic{equation}}

\section*{Water-network crossover interpolation}

The water-only largest-component fraction was evaluated on the nine principal
compositions and on an additional validation composition with 3,000 ethanol
molecules. Not-a-knot cubic-spline interpolation of the nine-point series
places the maximum-decline location at \(x^*=0.4335\). Including the
validation point, \(x_{\mathrm{EtOH}}=0.447962\), with measured
\(f_{\max}^{\mathrm{water}}=0.580446\), shifts the estimate to
\(x^*=0.4759\). These interpolated locations are operational finite-cell
descriptors, not directly sampled transitions.

\begin{figure}[!htbp]
\centering
\begin{minipage}[t]{0.48\textwidth}
\centering
\textbf{(a)}\par\smallskip
\includegraphics[width=\linewidth]{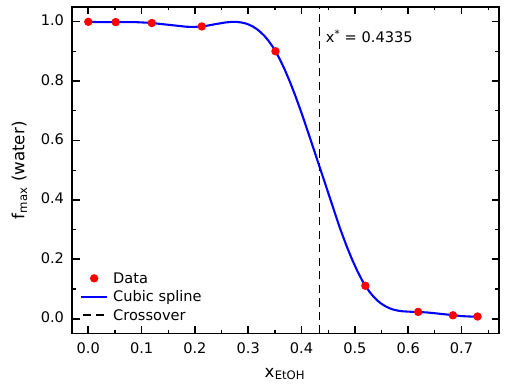}
\end{minipage}\hfill
\begin{minipage}[t]{0.48\textwidth}
\centering
\textbf{(b)}\par\smallskip
\includegraphics[width=\linewidth]{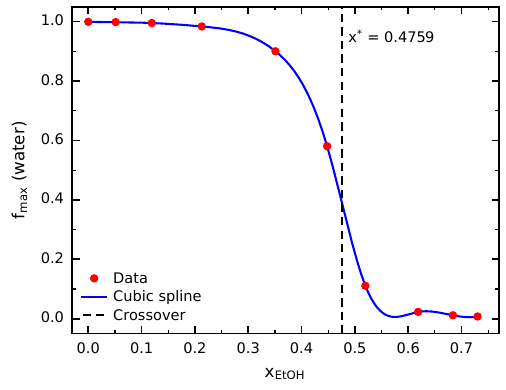}
\end{minipage}
\caption{Water-only largest-component fraction and cubic-spline
interpolation. (a) Fit to the nine principal compositions, with the
maximum-decline point at \(x^*=0.4335\). (b) Fit including the additional
3,000-ethanol validation composition at
\(x_{\mathrm{EtOH}}=0.447962\), where
\(f_{\max}^{\mathrm{water}}=0.580446\); its maximum-decline point is
\(x^*=0.4759\). These are operational finite-cell descriptors, not directly
sampled transitions.}
\label{fig:si-spline-fragmentation}
\end{figure}
\FloatBarrier

\subsection*{Crossover-estimate sensitivity to angular cutoff}

Angular-cutoff sensitivity was evaluated on the ten-composition grid,
including the 3,000-ethanol system at \(x_{\mathrm{EtOH}}=0.447962\), using
not-a-knot cubic-spline interpolation of the deduplicated 501-frame dataset.
The spline maximum-decline compositions are \(x_{\mathrm{EtOH}}=0.461\),
0.476, and 0.480 for cutoffs of 30, 35, and 40 degrees, respectively. The
total spread is \(\Delta x_{\mathrm{EtOH}}=0.020\) (0.0196 before rounding),
placing the operational fragmentation crossover near
\(x_{\mathrm{EtOH}}\approx0.46\text{--}0.48\) across the tested
30--40-degree range. This cutoff sensitivity is not a statistical uncertainty
and does not account for trajectory sampling, composition-grid resolution,
interpolation choices, or finite cell size.

\begin{table}[!htbp]
\centering
\caption{Angular sensitivity of the maximum-decline composition on the
ten-composition, deduplicated 501-frame dataset using not-a-knot cubic-spline
interpolation. The spread is not a statistical uncertainty.}
\label{tab:si-angular-sensitivity}
\begin{tabular}{ccc}
\toprule
Angular cutoff &  \(x^*\)  & \(s'(x^*)\)\\
\midrule
30$^\circ$ & 0.4608 & $-5.927$\\
35$^\circ$ & 0.4759 & $-7.236$\\
40$^\circ$ & 0.4804 & $-7.563$\\
\bottomrule
\end{tabular}
\end{table}
\FloatBarrier

\subsection*{Pure-water angular calibration and distance--angle distributions}

The following calibration is separate from the composition-resolved network
analysis above. It uses pure water with the standard TIP4P/2005
parameterization specified above (3,697 molecules) across 46 frames
(frames 0--450 in steps of 10), with the water--water distance boundary fixed
at 3.35~\AA{}. For this analysis, the donor-angle convention is
\(\theta_{\mathrm{HO_dO}}=\angle(\mathrm{H_d-O_d-O_a})\), whose vertex is
at the donor oxygen; a linear donor--hydrogen--acceptor arrangement is
therefore at zero degrees. The second angle,
\(\theta_{\mathrm{O_dHO_a}}=\angle(\mathrm{O_d-H-O_a})\), has its vertex
at hydrogen and is 180 degrees for a linear arrangement.

\begin{figure}[!htbp]
\centering
\begin{minipage}[t]{0.48\textwidth}
\centering
\textbf{(a) Candidate fraction retained}\par\smallskip
\includegraphics[width=\linewidth]{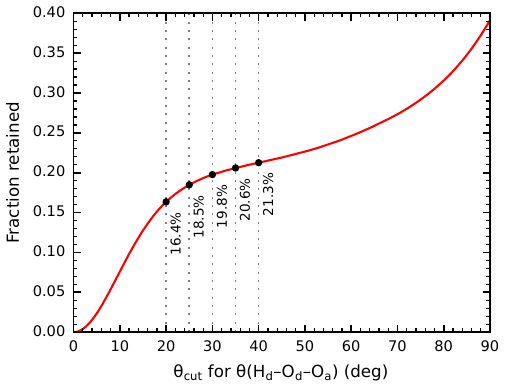}
\end{minipage}\hfill
\begin{minipage}[t]{0.48\textwidth}
\centering
\textbf{(b) Directed bonds and total participation}\par\smallskip
\includegraphics[width=\linewidth]{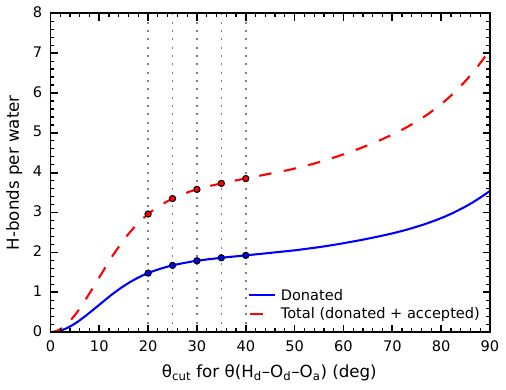}
\end{minipage}
\caption{Pure-water sensitivity to the donor-angle cutoff
\(\theta_{\mathrm{HO_dO}}\). (a) Fraction of directed candidates within the
strict 3.35~\AA{} distance boundary (1,541,300 candidates across 46 frames)
retained by the angular criterion. (b) Donated bonds per water and total
participation per water (donated plus accepted, equal to twice the donated
count). The dashed guides mark the tested cutoffs; 35$^\circ$ is the primary
criterion used in the main analysis. Curves are visual guides and no
uncertainty estimates are available.}
\label{fig:si-angle-cutoff-counts}
\end{figure}

The distance--angle maps use a 4.0~\AA{} candidate search radius
(2,658,948 directed donor-H to acceptor-O candidates) solely as a
visualization envelope for the joint distributions; this 4.0~\AA{} radius is
not used as the water-network cutoff or final hydrogen-bond cutoff. All
counted hydrogen bonds and marginal angular distributions are conditioned on
the stricter 3.35~\AA{} distance boundary (1,541,300 directed candidates).
Each passing geometry is one directed donor-H to acceptor-O hydrogen bond.
Thus, the fraction in Fig.~\ref{fig:si-angle-cutoff-counts}(a) is the
fraction of directed candidates within the 3.35~\AA{} distance boundary that
also satisfy the selected \(\theta_{\mathrm{HO_dO}}\) cutoff, while the
counts in panel (b) are expressed per water molecule over the 46 analyzed
frames. In pure water, mean accepted bonds per water equal mean donated bonds
per water, so total molecular participation equals twice the mean donated
count; for cluster and percolation analysis, the directed bond set is
collapsed into an undirected molecular graph only after both distance and
angle criteria are applied.

Raising the donor-angle cutoff from 20$^\circ$ to 40$^\circ$ increases the
directed count from 252,018 to 327,673 across these 46 frames (retaining
16.35\% to 21.26\% of the 1,541,300 directed candidates within 3.35~\AA{}).
The associated donated-bond population rises from 1.4819 to 1.9268 per water
(1.8665 at 35$^\circ$), and total donor-plus-acceptor participation rises
from 2.9638 to 3.8536 per water (3.7330 at 35$^\circ$). The increase is
monotonic over the tested cutoffs, but these single-system counts have no
uncertainty estimates. They quantify sensitivity of the operational geometric
definition; they do not independently validate the same cutoff for
ethanol-containing mixtures.

\begin{table}[!htbp]
\centering
\small
\caption{Pure-water hydrogen-bond counts at selected
\(\theta_{\mathrm{HO_dO}}\) cutoffs. The reference angle is the reported
\(\theta_{\mathrm{O_dHO_a}}\) value at the radial boundary; it is an observed
reference, not an exact geometrical complement. Candidate-retention fractions
are relative to the 1,541,300 directed donor-H to acceptor-O candidates
within the strict 3.35~\AA{} distance boundary.}
\label{tab:si-pure-water-angle-cutoff}
\begin{tabular}{cccccc}
\toprule
\(\theta_{\mathrm{HO_dO}}\) &
\(\theta_{\mathrm{O_dHO_a}}\) ref. & Directed bonds &
Candidate fraction & Donated per water & Participation per water \\
(degrees) & (degrees) & (46 frames) & retained & (mean) & (donated + accepted) \\
\midrule
20 & 152.391 & 252,018 & 0.1635 & 1.4819 & 2.9638 \\
25 & 145.745 & 284,929 & 0.1849 & 1.6754 & 3.3509 \\
30 & 139.251 & 304,580 & 0.1976 & 1.7910 & 3.5820 \\
35 & 132.923 & 317,418 & 0.2059 & 1.8665 & 3.7330 \\
40 & 126.768 & 327,673 & 0.2126 & 1.9268 & 3.8536 \\
\bottomrule
\end{tabular}
\end{table}

\begin{figure}[!htbp]
\centering
\begin{minipage}[t]{0.48\textwidth}
\centering
\textbf{(a) Donor-oxygen angle}\par\smallskip
\includegraphics[width=\linewidth]{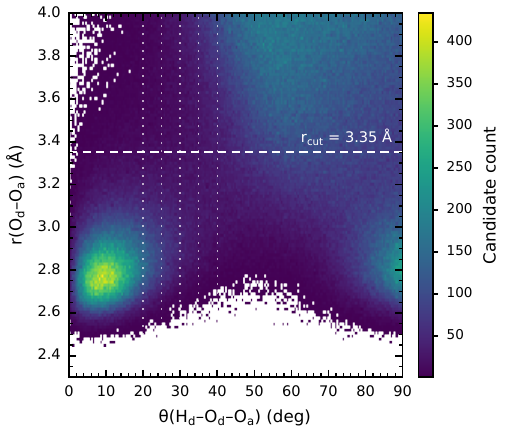}
\end{minipage}\hfill
\begin{minipage}[t]{0.48\textwidth}
\centering
\textbf{(b) Donor-hydrogen angle}\par\smallskip
\includegraphics[width=\linewidth]{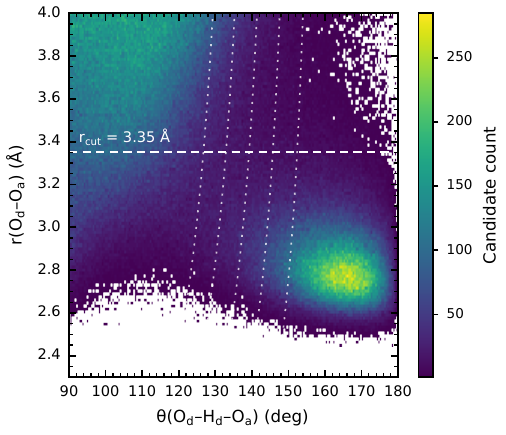}
\end{minipage}
\caption{Joint distance--angle distributions for pure-water candidate
geometries within the 4~\AA{} search radius. Panel (a) uses
\(\theta_{\mathrm{HO_dO}}\), with zero degrees corresponding to a linear
hydrogen bond; panel (b) uses \(\theta_{\mathrm{O_dHO_a}}\), with 180 degrees
corresponding to a linear bond. The dashed horizontal line marks
\(r_{\mathrm{cut}}=3.35\)~\AA{}, and vertical guides mark the tested angular
cutoffs. These maps visualize candidate geometry and are not themselves
the final bond-count criterion.}
\label{fig:si-angle-joint}
\end{figure}

\begin{figure}[!htbp]
\centering
\begin{minipage}[t]{0.48\textwidth}
\centering
\textbf{(a) Donor-oxygen angle}\par\smallskip
\includegraphics[width=\linewidth]{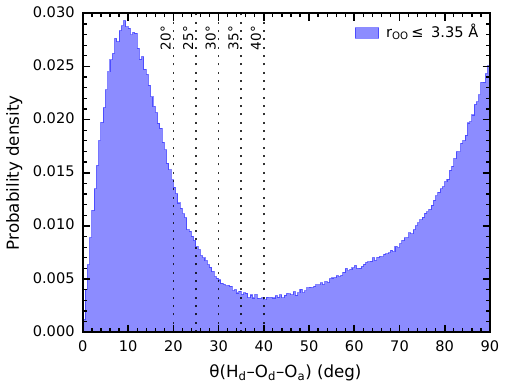}
\end{minipage}\hfill
\begin{minipage}[t]{0.48\textwidth}
\centering
\textbf{(b) Donor-hydrogen angle}\par\smallskip
\includegraphics[width=\linewidth]{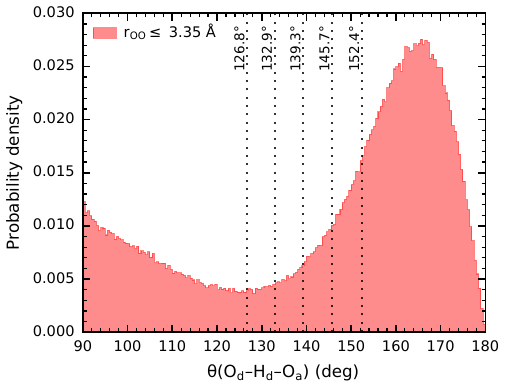}
\end{minipage}
\caption{Marginal angular distributions for pure-water geometries within
the strict radial boundary \(r\leq3.35\)~\AA{}. Panel (a) shows the
donor-oxygen angle used for the cutoff; panel (b) shows the donor-hydrogen
angle for the same geometries. Vertical guides in panel (a) indicate the
tested cutoffs, and corresponding reference angles are shown in panel (b).
The curves summarize the 46 analyzed frames and have no uncertainty bands.}
\label{fig:si-angle-marginal}
\end{figure}

Figures~\ref{fig:si-angle-joint} and~\ref{fig:si-angle-marginal}
resolve the geometry underlying this cutoff choice. The joint maps show
candidate counts as functions of donor--acceptor distance and each angle;
the horizontal line marks the 3.35~\AA{} radial criterion, and vertical
guides indicate selected donor-angle cutoffs. The marginal distributions
show the angular populations conditional on satisfying that radial
criterion. These distributions describe pure water only and are not
composition-specific validation data for the electrolyte systems.

\FloatBarrier

\section*{Temporal solvent- and ion-edge persistence}

This supplement documents the temporal edge-overlap analysis referenced in
the main article, including its definition, sampling details, and
composition-resolved solvent hydrogen-bond and ion-contact-edge curves.

\subsection*{Definition and sampling}

For an edge set \(E(t)\), endpoint-overlap persistence is
\begin{equation}
 C_{\mathrm{edge}}(\tau)=\left\langle
 \frac{|E(t)\cap E(t+\tau)|}{|E(t)|}\right\rangle_t.
 \label{eq:si-edge-persistence}
\end{equation}
Solvent edges are hydrogen bonds defined above. Ion-network edges are
Na--Cl contacts within the composition-specific first-shell boundary; they
are contact edges, not hydrogen bonds. This endpoint-overlap measure allows
an edge to break and reform between sampled observations; it is not a
continuous-survival probability. Curves cover 100 lags from 10 to 1,000~ps
in 10-ps increments for five compositions: 1,000, 2,000, 4,000, 6,000,
and 8,000 ethanol molecules. For each
edge class, the curve is normalized to its value at the first sampled lag:
\begin{equation}
 \widetilde C_j(\tau)=\frac{C_j(\tau)}{C_j(10\,\mathrm{ps})},
 \qquad 10\leq\tau\leq1{,}000~\mathrm{ps}.
 \label{eq:si-normalized-persistence}
\end{equation}
Thus, the plotted curves have no measured zero-lag point. Their area from
10 to 1,000~ps is calculated by trapezoidal quadrature as a descriptive
finite-window persistence measure, not as a complete edge lifetime. No
block-level uncertainties were calculated for the curves or areas, so
composition differences are descriptive rather than statistical tests.

\subsection*{Composition trends}

Water, ethanol, and mixed-solvent edge-overlap curves all decay strongly
over the sampled window. At 1,000~ps, all solvent-edge ordinates are below
0.011. Between 4,000 and 6,000 ethanol molecules, the water-edge area
increases from 68.88 to 75.49~ps (9.6\%), the ethanol-edge area from
50.05 to 51.94~ps (3.8\%), and the mixed-edge area from 62.06 to
63.81~ps (2.8\%). The corresponding 1,000-ps endpoint ordinates for
mixed edges are 0.00505 and 0.00506, and for ethanol edges 0.00322 and
0.00303, at 4,000 and 6,000 ethanol molecules, respectively. These small
changes and slight endpoint reversals do not establish saturation or
statistically significant composition differences.
\begin{figure}[!htbp]
\centering
\begin{minipage}[t]{0.48\textwidth}
\centering
\textbf{(a)}\par\smallskip
\includegraphics[width=\linewidth,height=0.28\textheight,keepaspectratio]{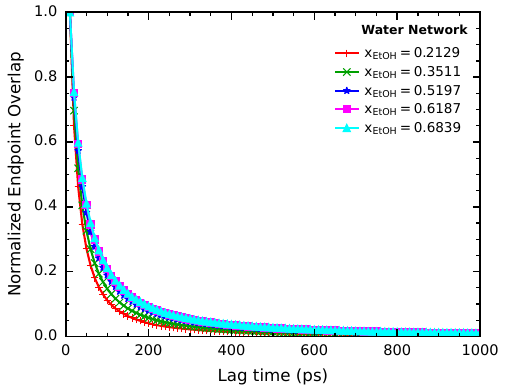}
\end{minipage}\hfill
\begin{minipage}[t]{0.48\textwidth}
\centering
\textbf{(b)}\par\smallskip
\includegraphics[width=\linewidth,height=0.28\textheight,keepaspectratio]{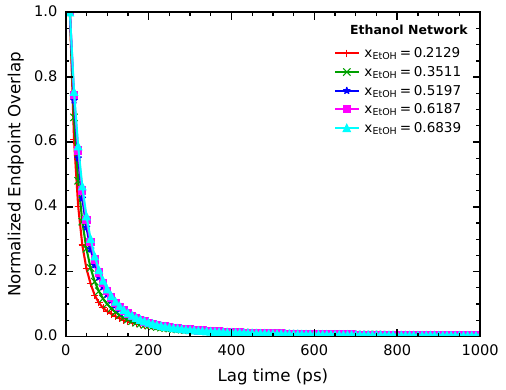}
\end{minipage}
\par\medskip
\begin{minipage}[t]{0.48\textwidth}
\centering
\textbf{(c)}\par\smallskip
\includegraphics[width=\linewidth,height=0.28\textheight,keepaspectratio]{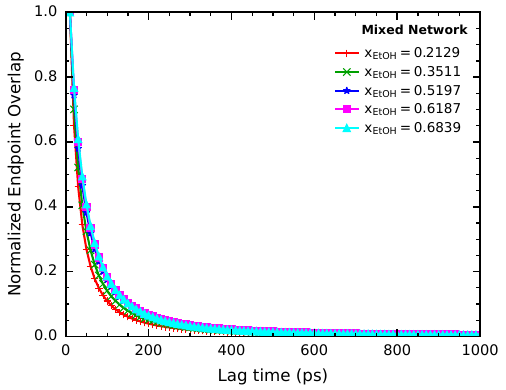}
\end{minipage}\hfill
\begin{minipage}[t]{0.48\textwidth}
\centering
\textbf{(d)}\par\smallskip
\includegraphics[width=\linewidth,height=0.28\textheight,keepaspectratio]{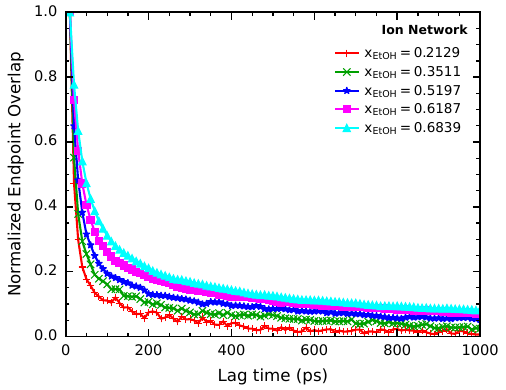}
\end{minipage}
\caption{Normalized endpoint-overlap persistence for (a) water, (b) ethanol,
(c) mixed-solvent, and (d) ion-network edges. Ion edges are Na--Cl contacts;
the solvent panels show hydrogen-bond edges. Every curve is normalized to its
value at 10~ps, the first sampled lag; no zero-lag point is shown. Lines
connect measured points and are not fitted relaxation models. Edge
reformation between observations can contribute to the curves.}
\label{fig:si-edge-persistence}
\end{figure}
\FloatBarrier

\section*{Electrostatic response methods}

\subsection*{Finite-field estimator, sampling, and uncertainty}

The solvent polarization is the sum of neutral water and ethanol molecular
dipoles; ionic charge displacements are excluded because they can reflect
sustained ionic current rather than bounded solvent polarization. The
fixed-volume estimator is
\begin{equation}
 M_{z,\mathrm{solv}}=\sum_{m\in\mathrm{water}}\mu_{m,z}
 +\sum_{m\in\mathrm{ethanol}}\mu_{m,z},
 \qquad
 \epsilon_s^{\mathrm{est}}=1+
 \frac{\langle M_{z,\mathrm{solv}}\rangle}{\epsilon_0 V_0 E_z},
 \label{eq:si-dielectric-estimator}
\end{equation}
where \(V_0\) is the fixed production-box volume of each composition, equal
to the final NPT volume (\(V_{\mathrm{prod}}\) in
Table~\ref{tab:si-volume-density}). This is a finite-field response, not a
zero-field static permittivity unless linear response is established.

At each composition, field-driven trajectories were run for 5~ns at
300~K in NVT using the final NPT box volume (Table~\ref{tab:si-volume-density}).
Uniform fields were applied along \(z\) at \(E_z=0.005\) and
\(0.007\ \mathrm{V/\text{\AA}}\). The initial 0.3~ns was omitted from
means and uncertainty estimates. Geyer initial-monotone-sequence
autocorrelation SEMs were computed from the retained solvent-dipole samples.
Augmented Dickey--Fuller tests over 0.5--5.0~ns reject the unit-root null at
all nine compositions for both fields (all \(p<1.1\times10^{-8}\)); this
test is not proof of strict stationarity. Composition-resolved test
statistics, selected lag counts, and observation counts are listed in
Tables~\ref{tab:si-adf-005} and~\ref{tab:si-adf-007}.

\begin{table}[!htbp]
\centering
\small
\caption{Augmented Dickey--Fuller test results for solvent-dipole samples
at \(E_z=0.005\ \mathrm{V/\text{\AA}}\), using the 0.5--5.0~ns window.
The null hypothesis is a unit root; the reported p-values test this null and
do not establish strict stationarity.}
\label{tab:si-adf-005}
\begin{tabular}{rccccc}
\toprule
\(N_{\mathrm{EtOH}}\) & \(x_{\mathrm{EtOH}}\) & ADF statistic &
\(p\)-value & Lags & Observations \\
\midrule
0 & 0.0000 & $-11.820$ & $8.46\times10^{-22}$ & 7 & 4,492 \\
200 & 0.0513 & $-12.592$ & $1.81\times10^{-23}$ & 2 & 4,497 \\
500 & 0.1191 & $-11.642$ & $2.13\times10^{-21}$ & 4 & 4,495 \\
1,000 & 0.2129 & $-10.208$ & $5.73\times10^{-18}$ & 3 & 4,496 \\
2,000 & 0.3511 & $-8.491$ & $1.31\times10^{-13}$ & 4 & 4,495 \\
4,000 & 0.5197 & $-7.298$ & $1.36\times10^{-10}$ & 4 & 4,495 \\
6,000 & 0.6187 & $-7.169$ & $2.83\times10^{-10}$ & 9 & 4,490 \\
8,000 & 0.6839 & $-6.895$ & $1.32\times10^{-9}$ & 8 & 4,491 \\
10,000 & 0.7301 & $-6.518$ & $1.06\times10^{-8}$ & 8 & 4,491 \\
\bottomrule
\end{tabular}
\end{table}

\begin{table}[!htbp]
\centering
\small
\caption{Augmented Dickey--Fuller test results for solvent-dipole samples
at \(E_z=0.007\ \mathrm{V/\text{\AA}}\), using the 0.5--5.0~ns window.
The null hypothesis is a unit root; the reported p-values test this null and
do not establish strict stationarity.}
\label{tab:si-adf-007}
\begin{tabular}{rccccc}
\toprule
\(N_{\mathrm{EtOH}}\) & \(x_{\mathrm{EtOH}}\) & ADF statistic &
\(p\)-value & Lags & Observations \\
\midrule
0 & 0.0000 & $-14.760$ & $2.41\times10^{-27}$ & 2 & 4,497 \\
200 & 0.0513 & $-11.704$ & $1.55\times10^{-21}$ & 3 & 4,496 \\
500 & 0.1191 & $-9.869$ & $4.05\times10^{-17}$ & 5 & 4,494 \\
1,000 & 0.2129 & $-8.623$ & $6.04\times10^{-14}$ & 5 & 4,494 \\
2,000 & 0.3511 & $-9.604$ & $1.89\times10^{-16}$ & 3 & 4,496 \\
4,000 & 0.5197 & $-7.278$ & $1.53\times10^{-10}$ & 14 & 4,485 \\
6,000 & 0.6187 & $-8.352$ & $2.97\times10^{-13}$ & 6 & 4,493 \\
8,000 & 0.6839 & $-9.583$ & $2.14\times10^{-16}$ & 6 & 4,493 \\
10,000 & 0.7301 & $-7.968$ & $2.83\times10^{-12}$ & 10 & 4,489 \\
\bottomrule
\end{tabular}
\end{table}
\FloatBarrier

\subsection*{Contact-ion index definition}

The structural index used in the main text is
\begin{equation}
 A_{\mathrm{CIP}}=\frac{f_{\mathrm{CIP}}}{f_{\mathrm{mono}}^2},
 \label{eq:si-cip-index}
\end{equation}
where \(f_{\mathrm{CIP}}\) is the fraction of ions with at least one Na--Cl
contact and \(f_{\mathrm{mono}}\) is the fraction with no Na--Cl contacts.
The first-shell Na--Cl boundaries span 3.35--3.48~\AA{}. Ions in larger
contact-connected aggregates contribute to \(f_{\mathrm{CIP}}\); the index
is not restricted to isolated dimers and is not a thermodynamic equilibrium
constant. The structural index \(A_{\mathrm{CIP}}\) was calculated from separate,
field-free 5-ns NVT trajectories sampled every 1~ps. Differencing successive
cumulative ion-cluster histograms yields a 1-ps series for the contact
fraction; Geyer-based autocorrelation SEMs were used where converged. At the
two highest ethanol loadings, the 6,000- and 10,000-ethanol single-trajectory
uncertainties did not converge, so final SEMs use four independent
trajectories: the original 5-ns run plus three 2-ns replicas
(\texttt{rep1}, \texttt{rep2}, and \texttt{rep3}). The
between-trajectory SEM is calculated as \(s_{\{\mu_i\}}/\sqrt{4}\) and
propagated to \(\ln A_{\mathrm{CIP}}\) by the delta method, giving relative
SEMs of 13.8\% and 7.8\% at 6,000 and 10,000 ethanol, respectively. With
only four trajectories of unequal length, these relative SEMs are themselves
approximate. These association and dielectric data are not from matched
trajectories and are compared only as composition-resolved observables.
\pagebreak

\subsection*{Volume and density summary}

Table~\ref{tab:si-volume-density} reports final production-box volume and
density alongside the NPT trajectory mean and SEM. In the supplied data,
the production entries equal the reported final NPT volume and density;
they are not an independent time average. The fixed-volume NVT runs,
including the finite-field runs, used the production-box (final NPT) volume;
\(V_0\) in Eq.~\eqref{eq:si-dielectric-estimator} is this value. The NPT mean
values are listed separately to make the distinction explicit; they differ
from the production values by up to about 0.7\% (at 200 ethanol).

\begin{table}[!htbp]
\centering
\small
\caption{Final production-box volume and density and corresponding NPT
trajectory means (mean~$\pm$~SEM). The production entries match the final
NPT values in the supplied summary; they are not separate production-run
averages.}
\label{tab:si-volume-density}
\begin{tabular}{rrrccr}
\toprule
$N_{\mathrm{EtOH}}$ & $x_{\mathrm{EtOH}}$ & $V_{\mathrm{prod}}$ (\AA$^3$) & $\rho_{\mathrm{prod}}$ (g\,cm$^{-3}$) & $\langle V\rangle_{\mathrm{NPT}}$ (\AA$^3$) & $\langle\rho\rangle_{\mathrm{NPT}}$ (g\,cm$^{-3}$) \\
\midrule
0 & 0.0000 & 112,906.9 & 1.0371 & $113,096.0\pm13.0$ & $1.0354\pm0.0001$ \\
200 & 0.0513 & 130,306.3 & 1.0227 & $131,213.3\pm51.8$ & $1.0157\pm0.0004$ \\
500 & 0.1191 & 158,369.0 & 0.9950 & $158,560.5\pm61.1$ & $0.9938\pm0.0004$ \\
1,000 & 0.2129 & 205,617.3 & 0.9632 & $205,339.4\pm73.8$ & $0.9645\pm0.0003$ \\
2,000 & 0.3511 & 301,880.3 & 0.9243 & $300,871.8\pm131.9$ & $0.9274\pm0.0004$ \\
3,000 & 0.4480 & 396,935.1 & 0.9069 & $397,929.7\pm160.4$ & $0.9046\pm0.0004$ \\
4,000 & 0.5197 & 493,833.4 & 0.8929 & $495,336.1\pm105.7$ & $0.8902\pm0.0002$ \\
6,000 & 0.6187 & 690,918.9 & 0.8726 & $691,031.0\pm206.1$ & $0.8724\pm0.0003$ \\
8,000 & 0.6839 & 888,793.2 & 0.8605 & $887,817.5\pm252.5$ & $0.8614\pm0.0002$ \\
10,000 & 0.7301 & 1,081,573.4 & 0.8568 & $1,084,454.8\pm393.5$ & $0.8545\pm0.0003$ \\
\bottomrule
\end{tabular}
\end{table}
\FloatBarrier

\section*{Thermodynamic equilibration and production stability}

\subsection*{Compression and NPT equilibration}

\begin{figure}[!htbp]
\centering
\eqpanel{(a) 0 ethanol molecules, $x_{\mathrm{EtOH}}=0.0000$}{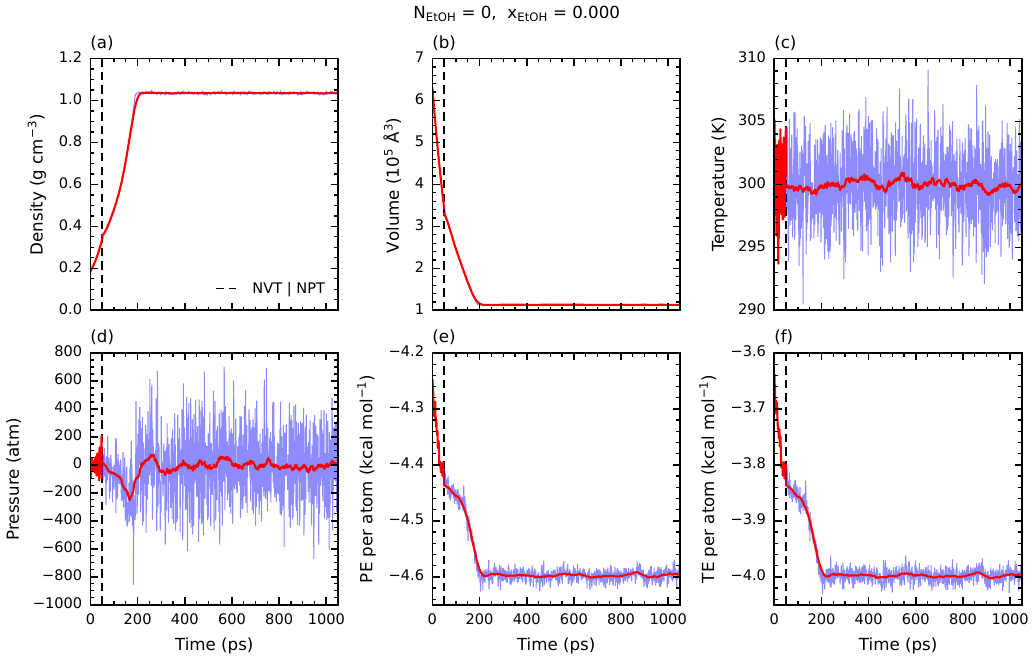}
\par\medskip
\eqpanel{(b) 200 ethanol molecules, $x_{\mathrm{EtOH}}=0.0513$}{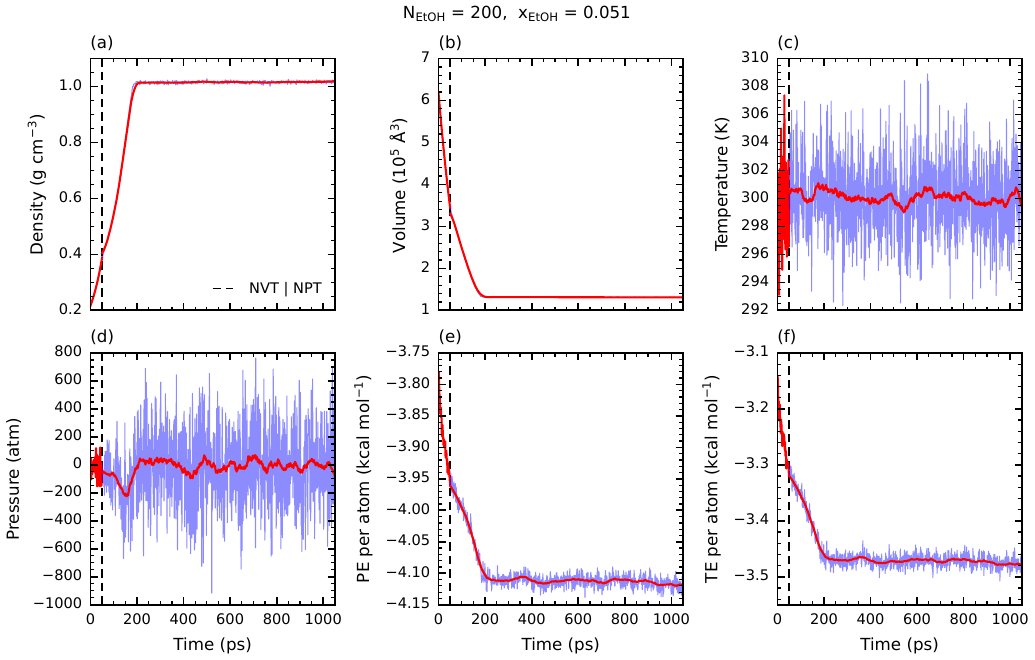}
\caption{Composition-resolved compression and NPT-equilibration traces
for (a) 0 and (b) 200 ethanol molecules. Panels (a)--(f) show density,
volume, temperature, pressure, potential energy per atom, and total energy per atom, respectively.}
\label{fig:si-equilibration-eth0-200}
\end{figure}

\begin{figure}[!htbp]
\centering
\eqpanel{(a) 500 ethanol molecules, $x_{\mathrm{EtOH}}=0.1191$}{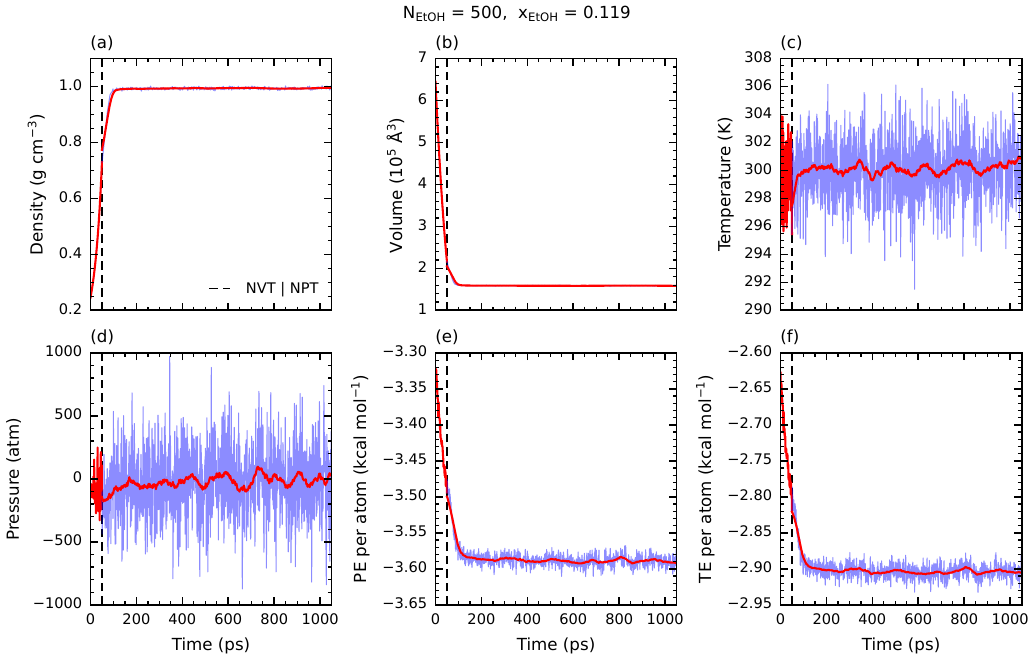}
\par\medskip
\eqpanel{(b) 1,000 ethanol molecules, $x_{\mathrm{EtOH}}=0.2129$}{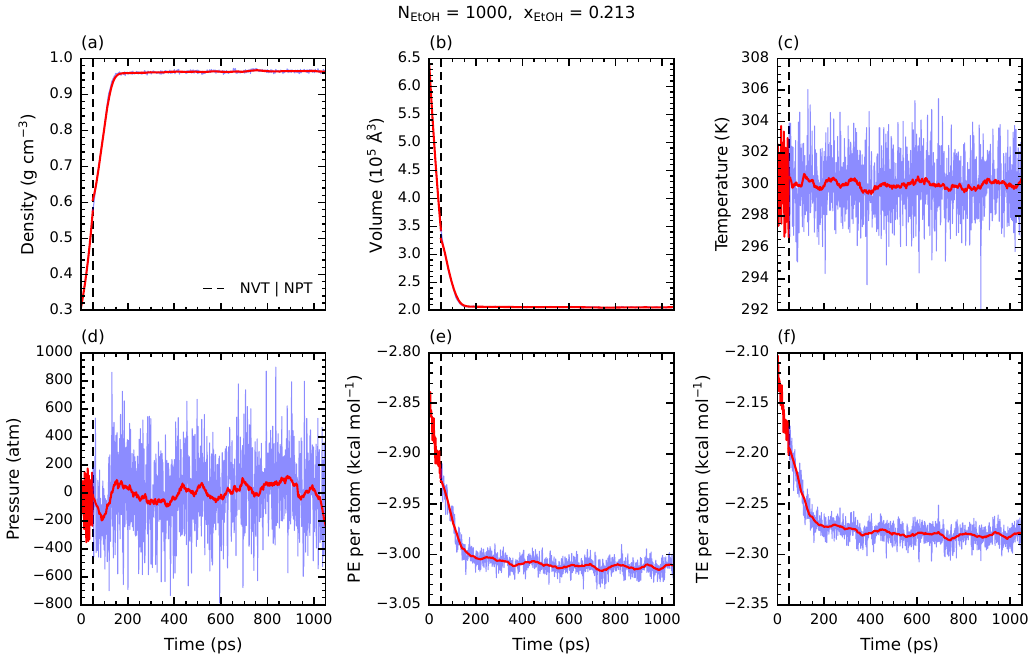}
\caption{Composition-resolved compression and NPT-equilibration traces
for (a) 500 and (b) 1,000 ethanol molecules. Panels (a)--(f) show density,
volume, temperature, pressure, potential energy per atom, and total energy per atom, respectively.}
\label{fig:si-equilibration-eth500-1000}
\end{figure}

\begin{figure}[!htbp]
\centering
\eqpanel{(a) 2,000 ethanol molecules, $x_{\mathrm{EtOH}}=0.3511$}{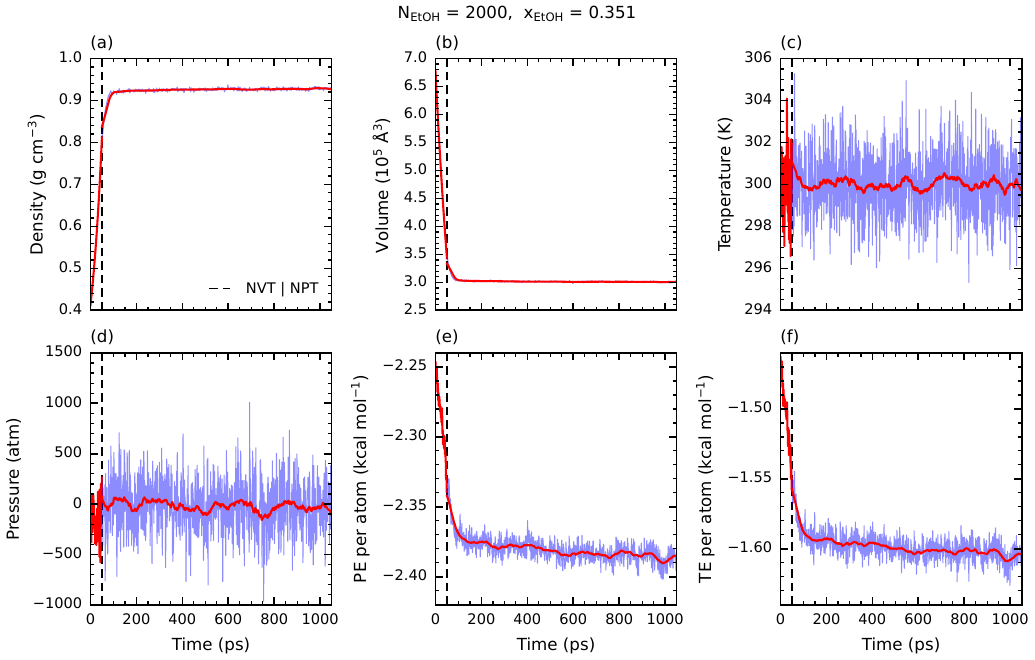}
\par\medskip
\eqpanel{(b) 3,000 ethanol molecules, $x_{\mathrm{EtOH}}=0.4480$}{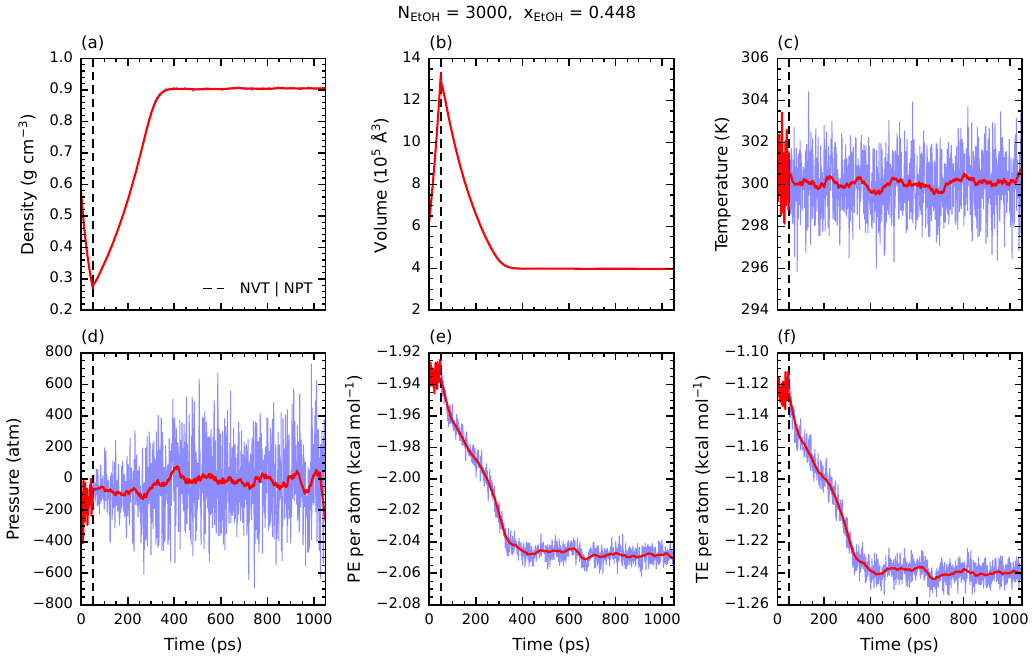}
\caption{Composition-resolved compression and NPT-equilibration traces
for (a) 2,000 and (b) 3,000 ethanol molecules. The 3,000-ethanol system
is the additional intermediate-composition validation system. Panels (a)--(f) show density,
volume, temperature, pressure, potential energy per atom, and total energy per atom, respectively.}
\label{fig:si-equilibration-eth2000-3000}
\end{figure}

\begin{figure}[!htbp]
\centering
\eqpanel{(a) 4,000 ethanol molecules, $x_{\mathrm{EtOH}}=0.5197$}{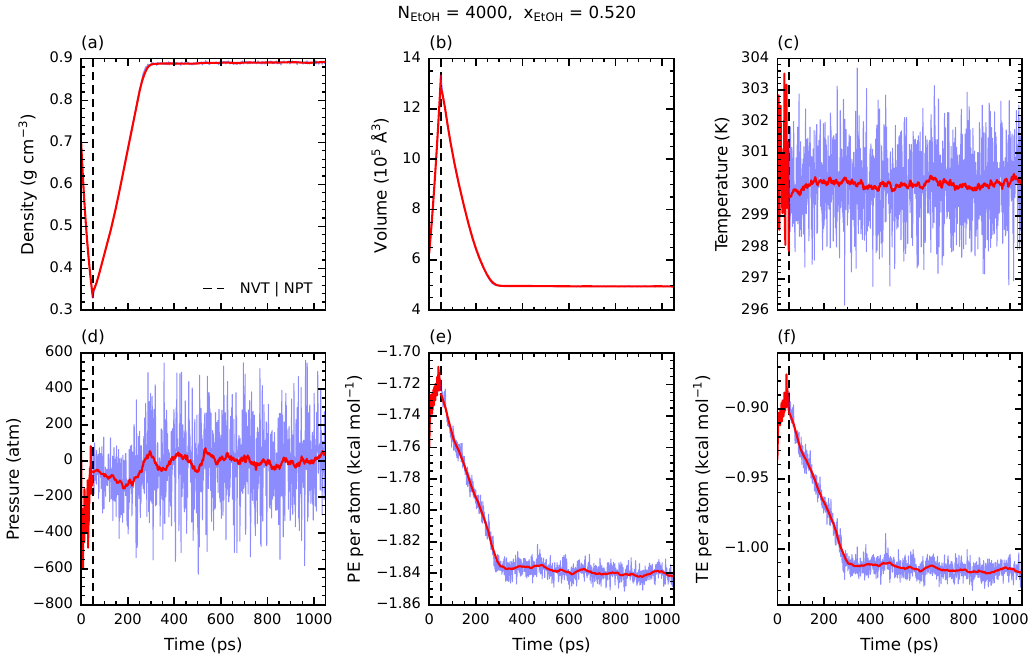}
\par\medskip
\eqpanel{(b) 6,000 ethanol molecules, $x_{\mathrm{EtOH}}=0.6187$}{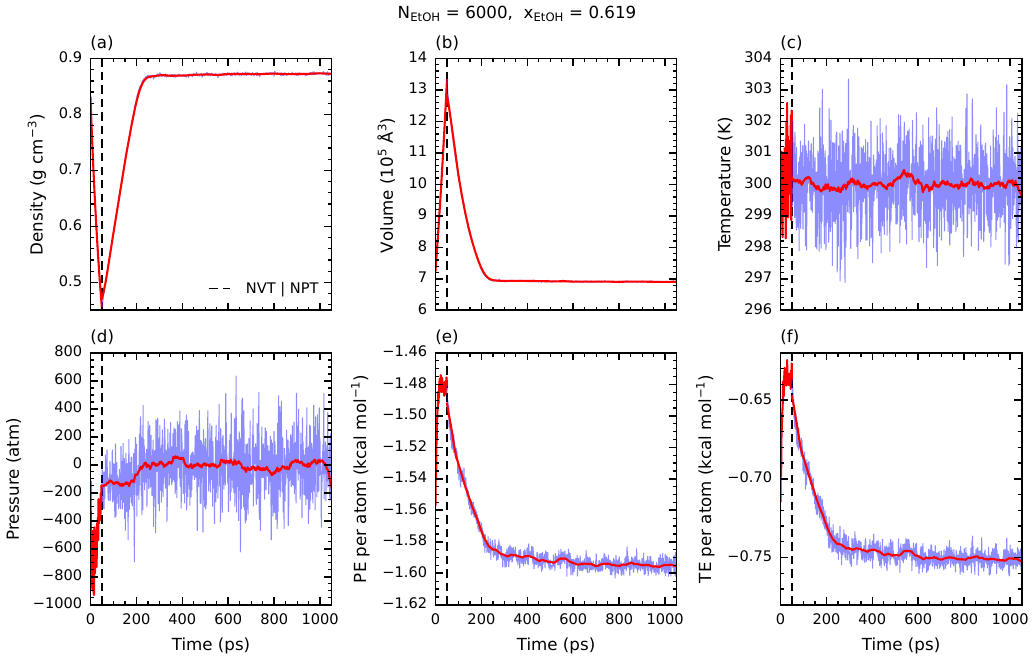}
\caption{Composition-resolved compression and NPT-equilibration traces
for (a) 4,000 and (b) 6,000 ethanol molecules. Panels (a)--(f) show density,
volume, temperature, pressure, potential energy per atom, and total energy per atom, respectively.}
\label{fig:si-equilibration-eth4000-6000}
\end{figure}

\begin{figure}[!htbp]
\centering
\eqpanel{(a) 8,000 ethanol molecules, $x_{\mathrm{EtOH}}=0.6839$}{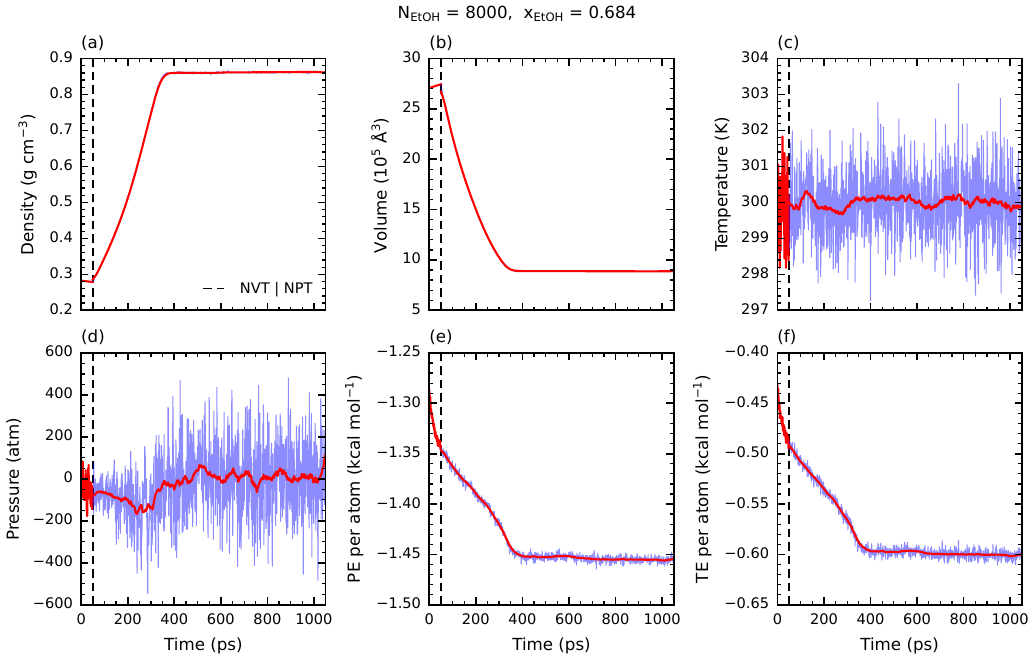}
\par\medskip
\eqpanel{(b) 10,000 ethanol molecules, $x_{\mathrm{EtOH}}=0.7301$}{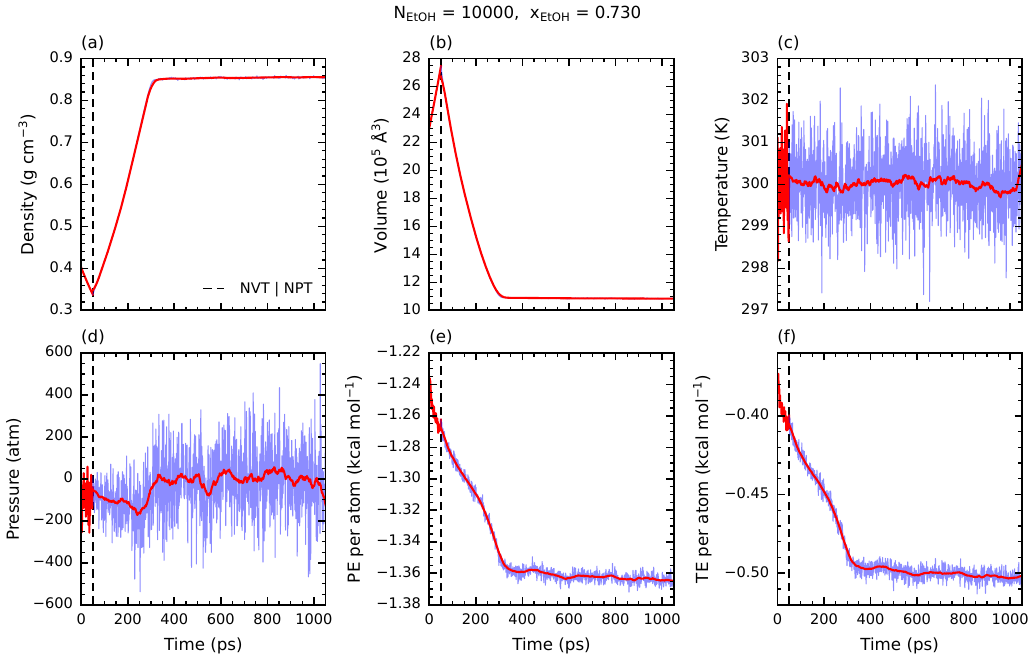}
\caption{Composition-resolved compression and NPT-equilibration traces
for (a) 8,000 and (b) 10,000 ethanol molecules. Panels (a)--(f) show density,
volume, temperature, pressure, potential energy per atom, and total energy per atom, respectively.}
\label{fig:si-equilibration-eth8000-10000}
\end{figure}

Composition-resolved traces show compression and NPT equilibration for all
ten systems (Figs.~\ref{fig:si-equilibration-eth0-200}--\ref{fig:si-equilibration-eth8000-10000});
dashed vertical markers denote the switch from initial NVT compression to NPT
equilibration. Density, volume, and per-atom energy traces show large initial
transients followed by flatter composition-specific profiles. Temperature
fluctuates around 300~K after its initial transient, while pressure remains
strongly fluctuating. These traces are useful relaxation and stability
diagnostics, but visual flattening alone does not establish equilibration or
quantify uncertainty in the resulting structural observables. Individual plots
at the end of this supplement resolve each composition; they are trajectory
diagnostics, not independent replicas. Mole fractions use the salt-free
water--ethanol solvent basis.

\subsection*{Production-run stability}

\begin{figure}[!htbp]
\centering
\includegraphics[width=\linewidth,height=0.5\textheight,keepaspectratio]{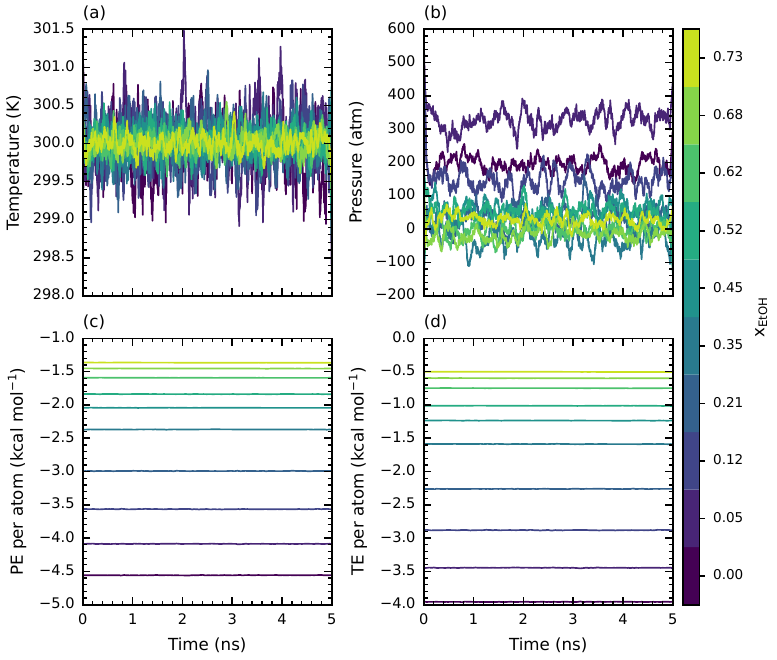}
\caption{Production-run stability diagnostics over 5~ns for all ten
compositions: (a) temperature, (b) pressure, (c) potential energy per
atom, and (d) total energy per atom. Color denotes solvent ethanol mole
fraction. The plotted traces are complemented by the mean, SEM, and
relative total-energy-drift summary in Table~\ref{tab:si-production-thermo}.}
\label{fig:si-production-stability}
\end{figure}

Figure~\ref{fig:si-production-stability} shows the production traces over
5~ns. Across the ten compositions, the tabulated mean temperatures lie
between 299.943 and 300.072~K. Pressure remains noisy;
the production-run mean pressures in Table~\ref{tab:si-production-thermo}
differ from the NPT values at several compositions (up to 327~atm at 200
ethanol); this offset is reported as a diagnostic, and its effect on the
finite-field estimate has not been quantified. Potential and total
energy traces appear comparatively flat at the plotted scale, while the
reported relative total-energy drift is nonzero and ranges from
$-0.14853$ to $+0.00462$\%~ns$^{-1}$, with the largest magnitude at the
10,000-ethanol composition. Thus the figure documents observed stability
over the analyzed window; it does not by itself establish equilibration or
zero energy drift.

\begin{table}[!htbp]
\centering
\setlength{\tabcolsep}{4pt}
\footnotesize
\caption{Thermodynamic summary by composition. Temperatures and pressures
are reported as mean~$\pm$~SEM; total-energy drift is the reported relative
drift per nanosecond. The pressure columns describe trajectory means and
should be interpreted with the large fluctuations visible in
Fig.~\ref{fig:si-production-stability}.}
\label{tab:si-production-thermo}
\begin{tabular}{rrcrrc}
\toprule
$N_{\mathrm{EtOH}}$ & $x_{\mathrm{EtOH}}$ & $T_{\mathrm{prod}}$ (K) &
$P_{\mathrm{NPT}}$ (atm) & $P_{\mathrm{prod}}$ (atm) &
Total-energy drift (\%~ns$^{-1}$) \\
\midrule
0 & 0.0000 & $299.943\pm0.052$ & $-0.2\pm9.0$ & $194.7\pm2.4$ & $+0.00462$ \\
200 & 0.0513 & $300.072\pm0.023$ & $0.5\pm6.7$ & $326.7\pm3.9$ & $-0.00171$ \\
500 & 0.1191 & $299.997\pm0.014$ & $-9.9\pm4.9$ & $145.0\pm5.4$ & $-0.00405$ \\
1,000 & 0.2129 & $299.987\pm0.013$ & $21.1\pm16.6$ & $33.9\pm4.4$ & $+0.00112$ \\
2,000 & 0.3511 & $300.004\pm0.020$ & $-29.3\pm9.2$ & $-25.1\pm4.4$ & $-0.01052$ \\
3,000 & 0.4480 & $300.052\pm0.022$ & $-13.1\pm6.1$ & $54.3\pm3.3$ & $+0.00037$ \\
4,000 & 0.5197 & $300.014\pm0.018$ & $-1.7\pm7.9$ & $50.6\pm3.7$ & $-0.03206$ \\
6,000 & 0.6187 & $299.994\pm0.009$ & $-4.5\pm12.8$ & $-3.2\pm1.9$ & $-0.06153$ \\
8,000 & 0.6839 & $300.007\pm0.013$ & $8.1\pm2.5$ & $-15.7\pm5.0$ & $-0.08970$ \\
10,000 & 0.7301 & $300.001\pm0.013$ & $3.3\pm5.3$ & $28.3\pm2.3$ & $-0.14853$ \\
\bottomrule
\end{tabular}
\end{table}

\clearpage

\clearpage

\clearpage

\clearpage

\clearpage

